\documentclass[american,english,aps,twocolumn,groupaddress,showpacs]{revtex4}
\usepackage[T1]{fontenc}
\usepackage[latin9]{inputenc}
\usepackage{amstext}
\usepackage{graphicx}
\usepackage{amssymb}

\makeatletter
\@ifundefined{textcolor}{}
{%
 \definecolor{BLACK}{gray}{0}
 \definecolor{WHITE}{gray}{1}
 \definecolor{RED}{rgb}{1,0,0}
 \definecolor{GREEN}{rgb}{0,1,0}
 \definecolor{BLUE}{rgb}{0,0,1}
 \definecolor{CYAN}{cmyk}{1,0,0,0}
 \definecolor{MAGENTA}{cmyk}{0,1,0,0}
 \definecolor{YELLOW}{cmyk}{0,0,1,0}
 }

\usepackage{times}
\renewcommand{\citet}{\cite}
\usepackage{mathptmx}
\usepackage{graphicx}
\usepackage{bm}

\makeatother

\usepackage{babel}

\begin{document}

\title{Heat conduction in graphene flakes with inhomogeneous mass interface}

\author{Jigger Cheh$^{1}$}

\email{jk_jigger@xmu.edu.cn}

\author{Hong Zhao$^{1,2}$}

\email{zhaoh@xmu.edu.cn}

\affiliation{$^{1}$Department of Physics, Institute of Theoretical Physics and
Astrophysics, Xiamen University, Xiamen 361005, China}

\affiliation{$^{2}$State Key Laboratory for Nonlinear Mechanics, Institute of
Mechanics, Chinese Academy of Sciences, Beijing 100080, China}

\pacs{65.80.Ck, 44.10.+i, 05.45.Yv}
\begin{abstract}
Using nonequilibrium molecular dynamics simulations, we study the
heat conduction in graphene flakes composed by two regions. One region
is mass-loaded and the other one is intact. It is found that the mass
interface between the two regions greatly decreases the thermal conductivity,
but it would not bring thermal rectification effect. The dependence
of thermal conductivity upon the heat flux and the mass difference
ratio are studied to confirm the generality of the result. The interfacial
scattering of solitons is studied to explain the absence of rectification
effect.
\end{abstract}
\maketitle
Thermal rectification is a phenomenon that the heat flux runs preferentially
along one direction and inferiorly along the opposite direction\citet{01.Casati.Nature,02.a review,03.Casati 2002,04.dio,05.zhangyong,06.deformed CNT,07.horns,08.Hu nanoletter,09.libaowen graphene,10.U-shaped,11.science,12.cnt1,13.cnt2,14.cnt3}.
It has attracted a great deal of attention in the last decade since
it reveals the possibility to control the heat transportation process.
With an improved understanding of thermal rectification, various devices
like thermal transistors, thermal logic circuits and thermal diodes
could be fabricated. Two methods are commonly used to design thermal
rectifiers. The first method is to couple two or more anharmonic chains
with different nonlinear potentials together\citet{03.Casati 2002,04.dio,05.zhangyong}.
The explanation for the observed rectification effect is that the
phonon bands of different regions of the chain will change from overlap
to separation when the heat flux is reversed. The asymmetry of interaction
potential controls the phonon band shift and it plays the central
role here. The second method is to implement asymmetric geometric
shape in quasi-1D and 2D systems. For example, it is applied in deformed
carbon nanotubes\citet{06.deformed CNT}, carbon nanohorns\citet{07.horns},
triangle shaped, trapezoid shaped and U-shaped graphene flakes\citet{08.Hu nanoletter,09.libaowen graphene,10.U-shaped}.
Thermal conductivity is higher when the heat flux runs from the narrow
to the wide region. The explanation for the observed rectification
is that the asymmetric geometric shape introduces asymmetric boundary
scattering of phonons. The asymmetry of geometric shape controls the
phonon scattering and it plays the central role in this case.

Recently a new procedure is considered by Chang et al. in carbon and
boron nitride nanotubes\citet{11.science}. They introduce the asymmetry
of mass distribution by covering external platinum compound on the
nanotubes. Comparing with nanotubes, the thermal contribution of the
platinum compound can be neglected. So the mass-loading procedure
is idealized by Chang et al. as changing the atomic mass of the atoms
in the heat conduction process. Higher thermal conductivity is observed
when the heat flux runs from the heavy mass region to the light mass
region. However the observed rectification effect cannot be explained
by phonons. First, the externally loaded molecules do not contribute
to any bond, thus the anharmonic coupling between the atoms in the
nanotubes is not changed. The asymmetry of interaction potential does
not contribute here. Second, the associated geometric deformation
of the nanotubes is not dominant, thus the boundary scattering of
phonons is also not changed. The asymmetry of geometric shape also
does not contribute in this case. Then Chang et al. surmise that the
origin of the rectification effect is due to the asymmetric interfacial
scattering between the two regions with inhomogeneous mass. However
as pointed by Change et al, it is well known that the interfacial
scattering of phonons is independent on the incident direction. Therefore
they postulate that solitons are responsible for the asymmetric interfacial
scattering process. Later similar thermal rectification effect has
been observed in molecular dynamic simulation of mass-graded carbon
nanotubes\citet{12.cnt1,13.cnt2,14.cnt3}. In the simulations, the
atomic mass of carbon atoms is gradually increased from 12 to 300
along the axis of tubes. The setup is treated as a combination of
multiple inhomogeneous mass interfaces. 

Graphene\citet{15.graphen1,16.graphen2}, a single layer of carbon
atoms in a honeycomb lattice with sp$^{2}$ bonds, reveals superior
high thermal conductivity up to 2500-5000 W/mK at room temperature\citet{17.graphene3,18.graphene4}.
It has been considered as a promising candidate for various thermal
devices. Graphene is similar to carbon nanotube in both structure
and thermal property, so it is interesting to know whether thermal
rectification would occur if the mass interface is implemented. Besides,
managing heat transportation across the interface is very common in
nanoelectronic design, thus it is also interesting to understand the
influence of mass interface upon the thermal conductivity of graphene. 

Here we study the heat conduction in graphene flakes with inhomogeneous
mass interface between two regions by NEMD (nonequilibrium molecular
dynamics) simulations. The graphene flakes are composed by two regions.
One region is intact and the other region is externally mass-loaded.
For simplicity the mass-loaded region is realized by modulating the
atomic mass of carbon atoms in the molecular dynamics simulations.
This setup is widely accepted as the idealized method for the mass-loading
procedure\citet{12.cnt1,13.cnt2,14.cnt3}. We report that thermal
rectification effect do not exist even a large mass difference ratio
is implemented. In order to understand the microscopic mechanism leading
to the absence of thermal rectification effect, we also study the
interfacial scattering of solitons in graphene.%
\begin{figure}
\includegraphics[scale=0.22]{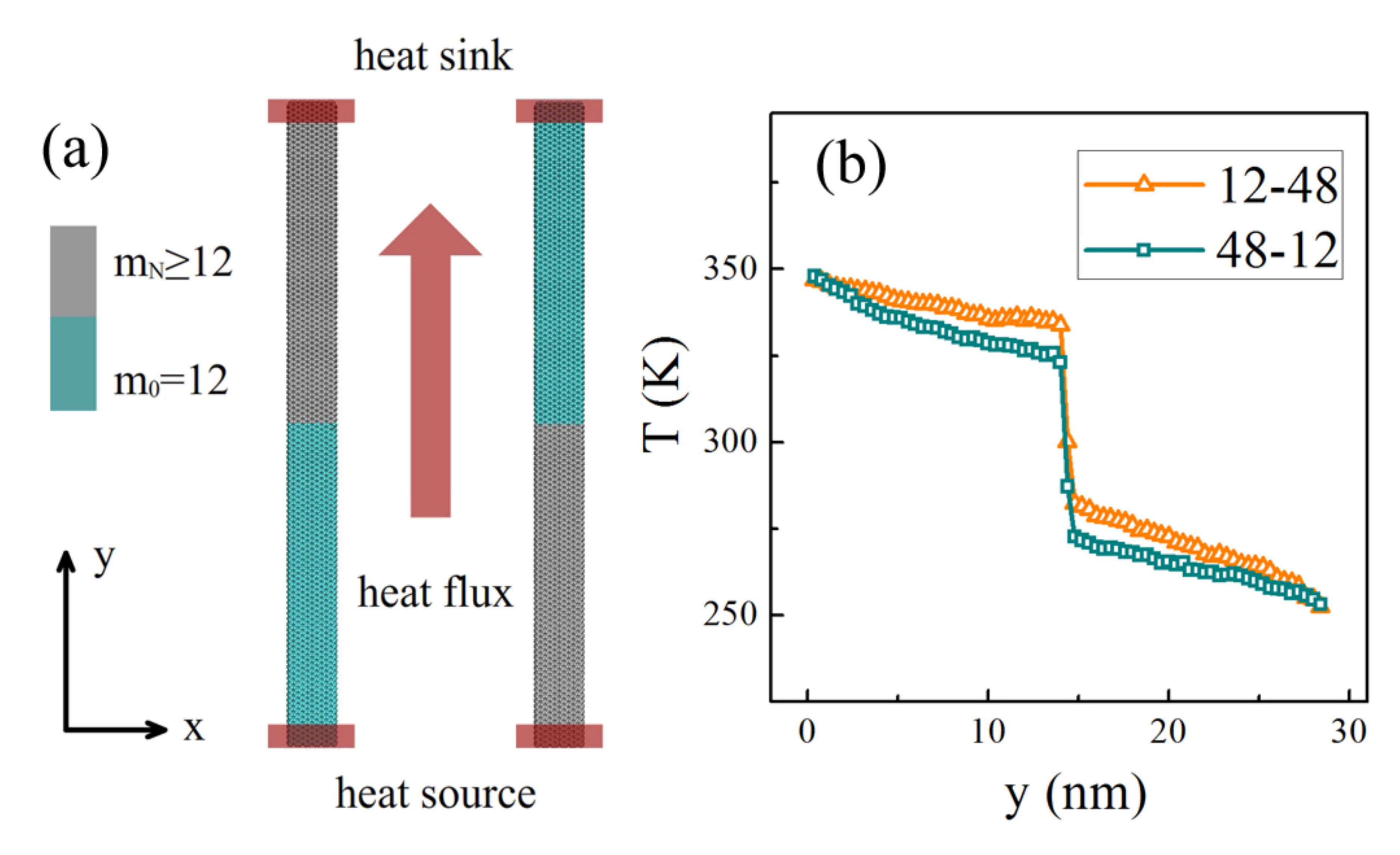}\caption{(a) Schematic of the graphene flakes with inhomogeneous mass interface.
The first one is labeled as the $m_{0}-m_{N}$ graphene flake. The
upper region of the $m_{0}-m_{N}$ graphene flake is mass-loaded which
is simplified by modulating the atomic mass of the carbon atoms as
$m_{N}\geqslant12$. Similarly the second one is labeled as the $m_{N}-m_{0}$
graphene flake. The lower region of the $m_{N}-m_{0}$ graphene flake
is mass-loaded. (b) The typical temperature profiles of the two graphene
flakes in the heat conduction process. Here $m_{N}/m_{0}=4$ and $J=0.35$
eV/ps are implemented. The $12-48$ and $48-12$ graphene flakes exhibit
the same temperature drop near the mass interface and between the
two ends along the whole simulation cell.}

\end{figure}

We carry out the molecular dynamics simulations of the heat conduction
in two rectangle graphene flakes with zig-zag edges along the x-axis
and armchair edges along the y-axis. The graphene flakes are shown
in Fig. 1(a). They are both 288 \foreignlanguage{american}{$\textrm{\AA}$}
long and 21 \foreignlanguage{american}{$\textrm{\AA}$} wide. The
heat sources and heat sinks are covered by red boxes with 100 carbon
atoms. The outmost edges of the heat sources/sinks are frozen. It
corresponds to fixed boundary conditions in the y-axis. Periodic boundary
condition is used along the x-axis. The atomic mass of the intact
carbon atoms is $m_{0}=12$ and they are drawn in cyan. The atomic
mass of the mass-loaded carbon atoms is $m_{N}\geqslant12$ and they
are drawn in silver. The upper half region of the first graphene flake
in Fig. 1(a) is mass-loaded, thus we label it as the as the $m_{0}-m_{N}$
graphene flake. Similarly we label the second one in Fig. 1(a) as
the $m_{N}-m_{0}$ graphene flake. The heat flux runs along the $m_{N}-m_{0}$
graphene flake is equivalent to the reversed heat flux runs along
the $m_{0}-m_{N}$ graphene flake. Thermal conduction process is investigated
by imposing the same heat flux along the two graphene flakes. It is
much more convenient later to compare the temperature profiles since
the heat sources and heat sinks are in the same direction. We use
the same reactive empirical bond-order (REBO) potential\citet{19.rebo}
as implemented in the LAMMPS\citet{20.lammps} code to simulate the
anharmonic coupling between the carbon atoms. Equations of motions
are integrated with velocity Verlet algorithm with the minimum timestep
$\bigtriangleup t=0.25$ fs.

First the graphene flakes are equilibrated at a constant temperature
T=300 K in the Nose-Hoover thermostat by 0.75 ns. After that the heat
flux is imposed. It is realized by the energy and momentum conserving
velocity rescaling algorithm developed by Jude and Jullien\citet{21.Jund}.
A constant rate of kinetic energy $dE$ is added in the heat source
and removed in the heat sink at each time step $dt$. The heat flux
is calculated as $J=dE/dt$. It is widely used to study the interfacial
heat conduction in different materials\citet{22.j1,23.j2,24.fluid}.
We divide the graphene flakes by several 4 \foreignlanguage{american}{$\textrm{\AA}$}
long slabs along the y-axis to obtain the temperature profiles. Each
slab contains about 60-70 carbon atoms. The local temperature in each
slab is calculated from the averaged kinetic energy of the carbon
atoms. The temperature profiles are averaged over every 100 ps after
the heat flux is imposed. The whole nonequilibrium simulation process
covers 3 ns and the last 1 ns is utilized as the steady state since
the temperature profiles do not change much. Thermal conductivity
$G$ of the whole graphene flake is calculated by the Fourier\textquoteright{}s
Law: 

\begin{equation}
G=-\frac{J/A}{\bigtriangleup T/\bigtriangleup L}\end{equation}

\noindent Here $J$ is the heat flux, $A$ is the cross section of
the heat transfer defined by the product of the width and the thickness
of the graphene flakes. The thickness of the graphene flake is usually
considered as the bond length of the carbon atoms and it is taken
as 1.4 \foreignlanguage{american}{$\textrm{\AA}$} in our computation\citet{25.thickness1,26.thickness2}.
$\bigtriangleup T$ ($\bigtriangleup L$) is the temperature (distance)
difference between the two ends on the graphene flake. Thermal conductivity
G represents the thermal conducting capacity of the whole graphene
flake. eV/ps is selected as the unit for the heat flux, \foreignlanguage{american}{$\textrm{\AA}$}
is selected as the unit for the width and thickness, K is selected
as the unit for the temperature. W/mK is selected as the unit for
the thermal conductivity, so the associated numerical results are
converted to this unit. 

In Fig. 1(b) we show the typical temperature profiles of the two graphene
flakes. Here $m_{N}/m_{0}=4$ and $J=0.35$ eV/ps are implemented.
A sudden temperature drop is observed near the mass interface. It
indicates that the mass interface behaves like a strong thermal resistive
boundary. In Fig. 1(b), it also shows that although the temperature
profiles are different in the two graphene flakes, but the temperature
differences between the two ends are the same. The temperature difference
in the $12-48$ graphene flake is 94.0 K and the temperature difference
in the $48-12$ graphene flake is 93.8 K. Thus the thermal conductivity
G of the two graphene flakes are the same. It indicates that even
for such a large temperature difference, there is still no observable
thermal rectification effect. The $12-48$ and the $48-12$ graphene
flakes have the same thermal conducting capacity and the heat flux
runs equivalently without preferred direction. 

In order to further confirm the result that there is no thermal rectification
effect brought by the mass interface in the graphene flakes, the dependence
of thermal conductivity upon the heat flux and the mass difference
ratio are studied. First we keep the mass difference ratio $m_{N}/m_{0}=4$
invariant and vary the heat flux from 0.15 to 0.5 eV/ps. In Fig. 2(a)
we show the dependence of thermal conductivity upon the heat flux.
Thermal conductivity is almost the same in the two graphene flakes.
It indicates that there is no thermal rectification effect by using
different value of heat flux. Meanwhile it is found that thermal conductivity
is decreasing with the heat flux. For $J=0.15$ eV/ps, the thermal
conductivity is about 110 W/mK. For $J=0.5$ eV/ps, it is reduced
to 38 W/mK which is about 35\% of the original value. It indicates
that interfacial scattering is enhanced by the amount of heat flux
and the effect might be taken into consideration in real application. 

Second we keep the heat flux $J=0.35$ eV/ps invariant and vary the
mass difference ratio $m_{N}/m_{0}$ from 1 to 5. Here $m_{N}/m_{0}=1$
stands for the graphene flake without the mass interface. In Fig.
2(b) we show the dependence of thermal conductivity upon the mass
difference ratio. Thermal conductivity is almost the same in the two
graphene flakes. It indicates there is no thermal rectification effect
by using different mass difference ratio. Meanwhile it is found that
thermal conductivity is greatly decreased by the mass interface. The
thermal conductivity of the graphene flake without the mass interface
is about 269 W/mK. When the mass interface is implemented ($m_{N}/m_{0}=5$),
it is reduced to 45 W/mK. It is only about 16\% of the original value.
Here it provides a possible route to tune the thermal behavior of
graphene. The mass interface can be implemented by two methods. One
is to load external heavy and thermal insulating molecules upon the
carbon atoms\citet{11.science,12.cnt1,13.cnt2,14.cnt3}. The other
one is to use different ratio of isotope substitutions\citet{27.iso,28.iso,29.ios letter}
which is demonstrated possible in experiment by chemical vapor deposition
growth of graphene on metal\citet{30.iso}.%
\begin{figure}
\includegraphics[scale=0.2]{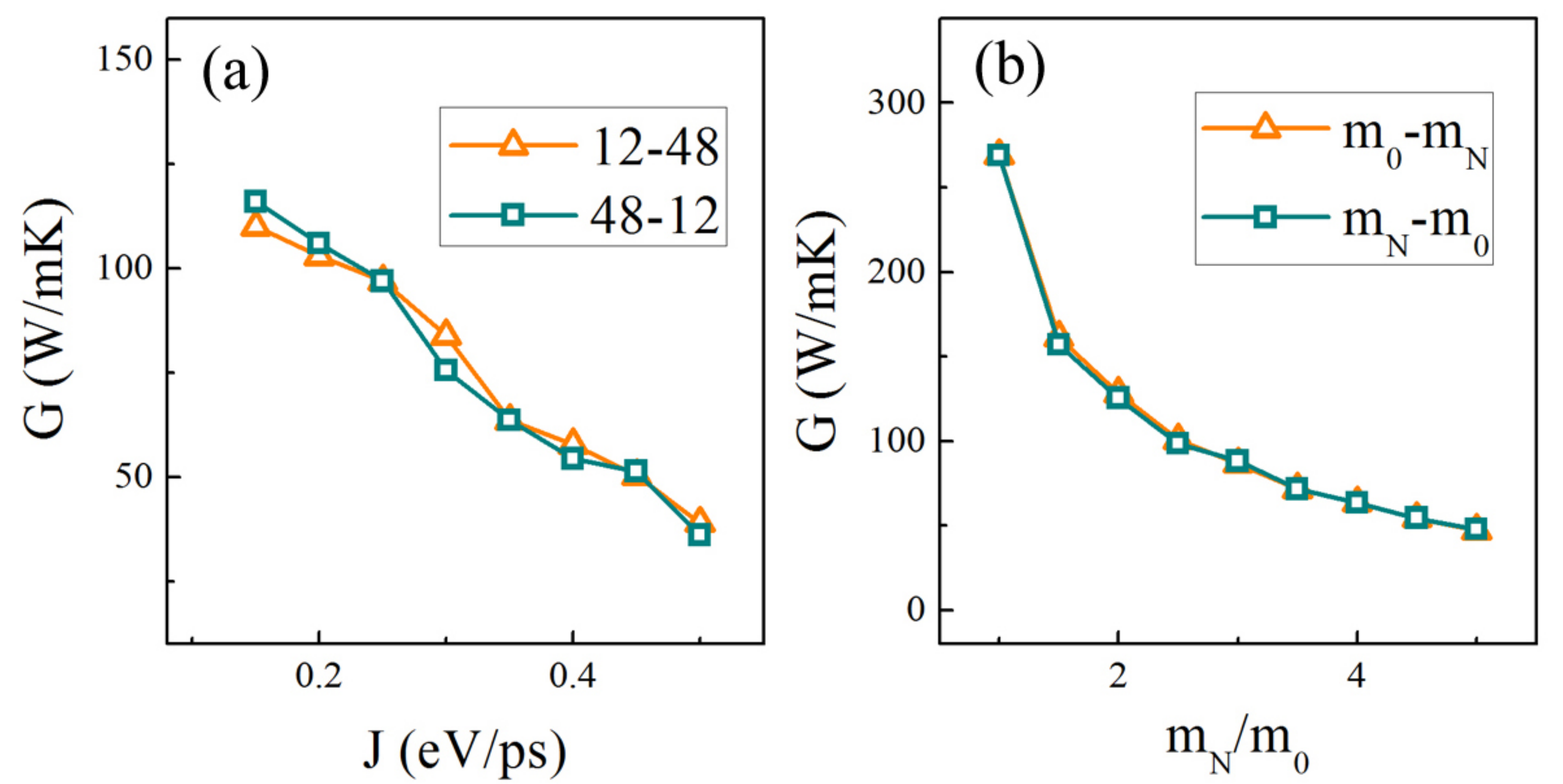}

\caption{(a) Thermal conductivity $G$ vs heat flux $J$. Here the mass difference
ratio $m_{N}/m_{0}=4$ is unchanged and the heat flux $J$ varies
from 0.15 to 0.5 eV/ps. (b) Thermal conductivity G vs the mass difference
ratio $m_{N}/m_{0}$. Here the heat flux $J=0.35$ eV/ps is unchanged
and the mass difference ratio varies from 1 to 5. For $m_{N}/m_{0}=1$,
the thermal conductivity $G$ is obtained from the graphene flake
without the mass interface.}

\end{figure}
\begin{figure}
\includegraphics[scale=0.2]{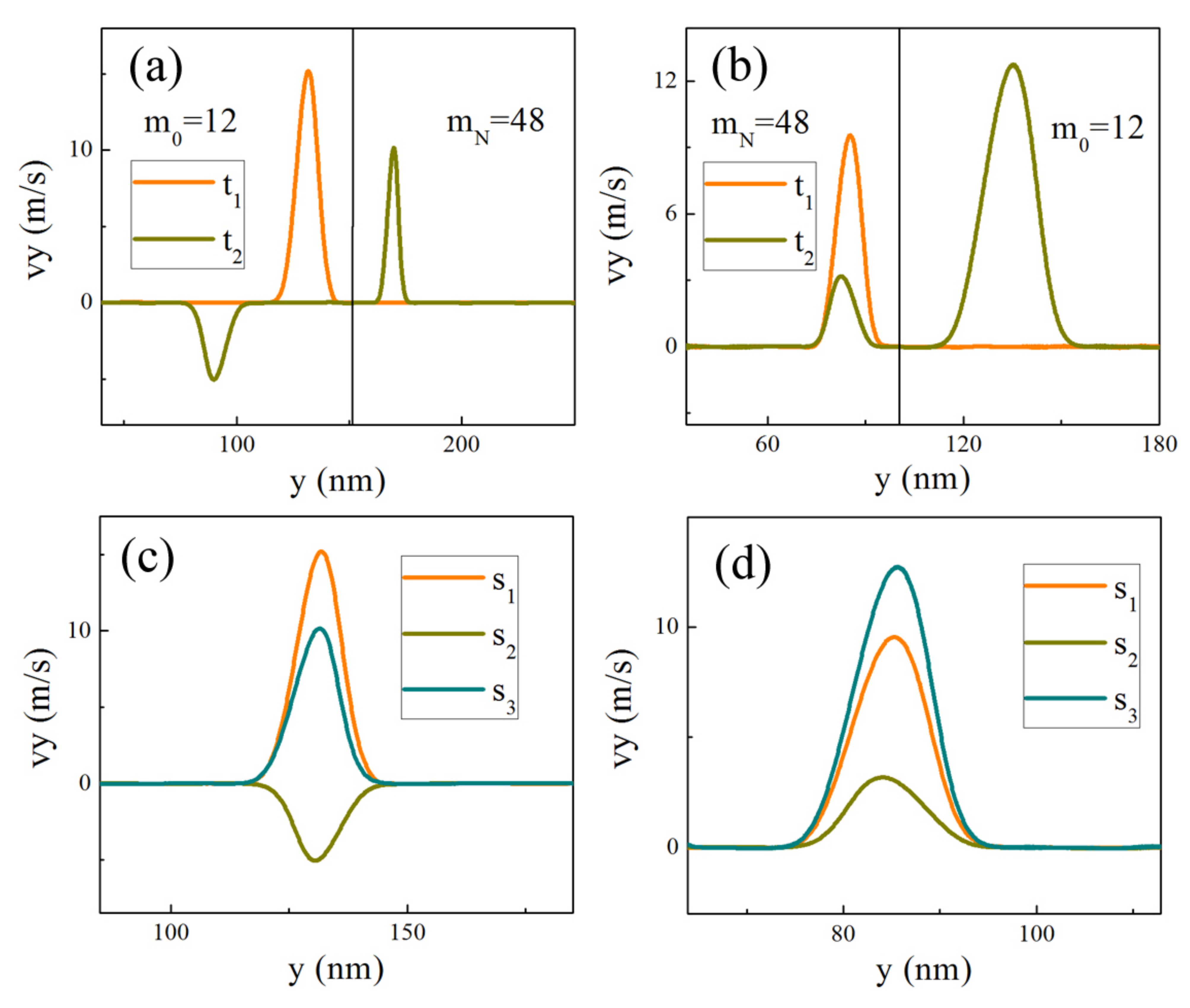}

\caption{{[}(a)-(b){]} The barrier line in the middle of (a) and (b) separates
the graphene flake into two regions. The atomic mass of the $m_{0}$
region is 12. The atomic mass of the $m_{N}$ region is 48. The curve
t$_{1}$ represents the longitudinal velocity of the carbon atoms
before the scattering. The curve t$_{2}$ represents the longitudinal
velocity of the carbon atoms after the scattering. The time interval
between t$_{1}$ and t$_{2}$ is 3.2 ps. The $12-48$ scattering is
shown in (a). The $48-12$ scattering is shown in (b). {[}(c)-(d){]}
Rescaling the width of the reflected soliton s$_{2}$ and the transmitted
soliton s$_{3}$ according to the width of the incident soliton s$_{1}$.
The rescaling in the $12-48$ scattering is shown in (c). The rescaling
in the $48-12$ scattering is shown in (d).}

\end{figure}

The heat conduction results suggest that inhomogeneous mass interface
cannot lead to thermal rectification effect in graphene. In the above
simulations, the anharmonic coupling between the carbon atoms is kept
constant and there is no geometric deformation by using periodic boundary
condition along the width. Therefore just like carbon nanotubes, the
asymmetry of interaction potential and the asymmetry of geometric
shape also do not contribute in the graphene flakes. Furthermore,
the interfacial scattering of solitons in surmised to be responsible
for the thermal rectification effect in carbon nanotubes\citet{11.science}.
So in order to understand the absence of thermal rectification effect,
it is necessary to study the interfacial scattering of solitons in
graphene. Recently, subsonic NLS (Nonlinear Schrödinger) equation
described solitons are found in graphene\citet{31.zhao Jigger}. They
preserve their identities in propagation and exhibit strong interactions
and phase shifts in collision. Their energy reflection rates between
the two regions with different mass are needed if one tends to know
whether the solitons would bring rectification effect in the interfacial
scattering process\citet{32.JPS,33.grn1,34.grn2}.

Here we first generate a longitudinal soliton in a 485 nm long graphene
flake with atomic mass $m_{0}=12$.The chirality of the graphene flake
is the same as the two graphene flakes in Fig. 1(a). In the curve
t$_{1}$ of Fig. 3(a), we show the longitudinal velocity (vy) distribution
of the carbon atoms before the interfacial scattering. The soliton
is set as the incident soliton and its amplitude is positive. After
that we change the atomic mass of the carbon atoms ahead of the solitons
to be $m_{N}=48$ to set up a mass interface. The barrier line in
the middle of Fig. 3(a) separates the graphene flake into two regions.
The atomic mass of the left region is intact ($m_{0}=12$) and the
atomic mass of the right region is mass-loaded ($m_{N}=48$). The
interfacial scattering from the $m_{0}$ to the $m_{N}$ region happens
when the incident soliton hits the mass interface. In the curve t$_{2}$
of Fig. 3(a), we show the longitudinal velocity distribution of the
carbon atoms after the interfacial scattering. The incident soliton
is scattered to a transmitted soliton and a reflected soliton. The
amplitude of the transmitted soliton is positive and the amplitude
of the reflected soliton is negative in the $12-48$ scattering. Similarly
we obtain the interfacial scattering from the $m_{N}$ to the $m_{0}$
region. In Fig. 3(b) we show that the amplitude of the transmitted
soliton is positive and the amplitude of the reflected soliton is
also positive in the $48-12$ scattering. The scattering results indicate
that the amplitude of the reflected soliton is dependent upon the
incident direction. It is negative in the $m_{0}-m_{N}$ scattering
and positive in the $m_{N}-m_{0}$ scattering. Such behavior of the
reflected soliton has also been obtained in the interfacial scattering
of similar NLS solitons in 1D nonlinear chain\citet{32.JPS}. It represents
a different kinetic behavior of the NLS solitons from the KdV solitons
which have no reflected solitons in the $m_{N}-m_{0}$ scattering.

There is a rescaling relation between the width of the scattered solitons.
The propagating velocity of a soliton in the $m_{0}$ region is 20
km/s and in the $m_{N}$ section\citet{31.zhao Jigger} is $v(m_{N})=[cos\frac{kl_{0}}{2}-(\alpha+\frac{Atan\varphi}{2\sqrt{\sigma}})sin\frac{kl_{0}}{2}]\sqrt{\frac{b}{m_{N}}}$$\propto1/\sqrt{m_{N}}$.
Since the scattering time $\bigtriangleup t$ is very short thus we
can neglect the amplitude dependent parameters in the formula to estimate
$\bigtriangleup t$. In the $m_{0}-m_{N}$ scattering $\bigtriangleup t$
can be estimated by the width of the incident soliton $w_{m_{0}-m_{N}}^{I}$
as $\triangle t=w_{m_{0}-m_{N}}^{I}/20$. So the width of the reflected
soliton $w_{m_{0}-m_{N}}^{R}$ in the $m_{0}$ region and the width
of transmitted soliton $w_{m_{0}-m_{N}}^{T}$ in the $m_{N}$ region
are:

\begin{equation}
w_{m_{0}-m_{N}}^{R}=20\triangle t=w_{m_{0}-m_{N}}^{I}\end{equation}

\begin{equation}
w_{m_{0}-m_{N}}^{T}=v(m_{N})\triangle t=\frac{\sqrt{m_{0}}}{\sqrt{m_{N}}}w_{m_{0}-m_{N}}^{I}\end{equation}

\noindent Similar relations can be obtained in $m_{N}-m_{0}$ scattering.
The width of the reflected soliton $w_{m_{N}-m_{0}}^{R}$ in the $m_{N}$
region and the width of transmitted soliton $w_{m_{N}-m_{0}}^{T}$
in the $m_{0}$ region are:

\begin{equation}
w_{m_{N}-m_{0}}^{R}=w_{m_{N}-m_{0}}^{I}\end{equation}

\begin{equation}
w_{m_{N}-m_{0}}^{T}=\frac{\sqrt{m_{N}}}{\sqrt{m_{0}}}w_{m_{N}-m_{0}}^{I}\end{equation}

\noindent In Fig. 3(c) and Fig. 3(d) we show the width rescaling relations
between the scattered solitons. The width of the reflected soliton
s$_{2}$ and the transmitted soliton s$_{3}$ are rescaled well according
to the width of the incident soliton s$_{1}$. 

In order to understand the role of solitons play in the heat conduction
process, the energy reflection rate is needed. We measure the energy
$E$ and the momentum $P$ of a soliton in graphene as the aggregated
momentum and energy of all the carbon atoms along the width of the
soliton\citet{35.zhao,36.zhao2}:

\begin{equation}
E=\sum_{k}\frac{1}{2}mv_{k}^{2}+E_{k}^{REBO}\qquad P=\sum_{k}mv_{k}\end{equation}

\noindent Here only the carbon atoms with $\left\Vert v_{k}\right\Vert >0.01$
m/s are counted. The widths of the scattered solitons also could be
estimated by the rescaling relation in Eq. (2)-(5). When the energy
and the momentum of the scattered solitons are obtained, the energy
reflection rate $R^{E}$ and the momentum reflection rate $R^{P}$
are defined as:

\begin{equation}
R^{E}=\frac{E^{R}}{E^{I}}\times100\%\qquad R^{P}=\frac{P^{R}}{P^{I}}\times100\%\end{equation}

\noindent Here $E^{I}$ and $P^{I}$ are the energy and the momentum
of the incident soliton. $E^{R}$ and $P^{R}$ are the energy and
the momentum of the reflected soliton.

In Fig. 4 we show the dependence of the energy and the momentum reflection
rate upon the mass difference ratio. In Fig. 4(a) it shows that the
energy reflection rates in the $m_{0}-m_{N}$ and the $m_{N}-m_{0}$
scattering are the same. In Fig. 4(b) it shows that the momentum reflection
rates are asymmetric. In the $m_{0}-m_{N}$ ($m_{N}-m_{0}$) scattering,
the negative (positive) momentum reflection rates are obtained. The
results explain the absence of thermal rectification in graphene:
although the momentum reflection rate is dependent upon the incident
direction, the energy reflection rate is directional independent.
The same amount of energy would be reflected by the mass interface,
thus it brings no reflection effect in the heat conduction process.%
\begin{figure}
\includegraphics[scale=0.25]{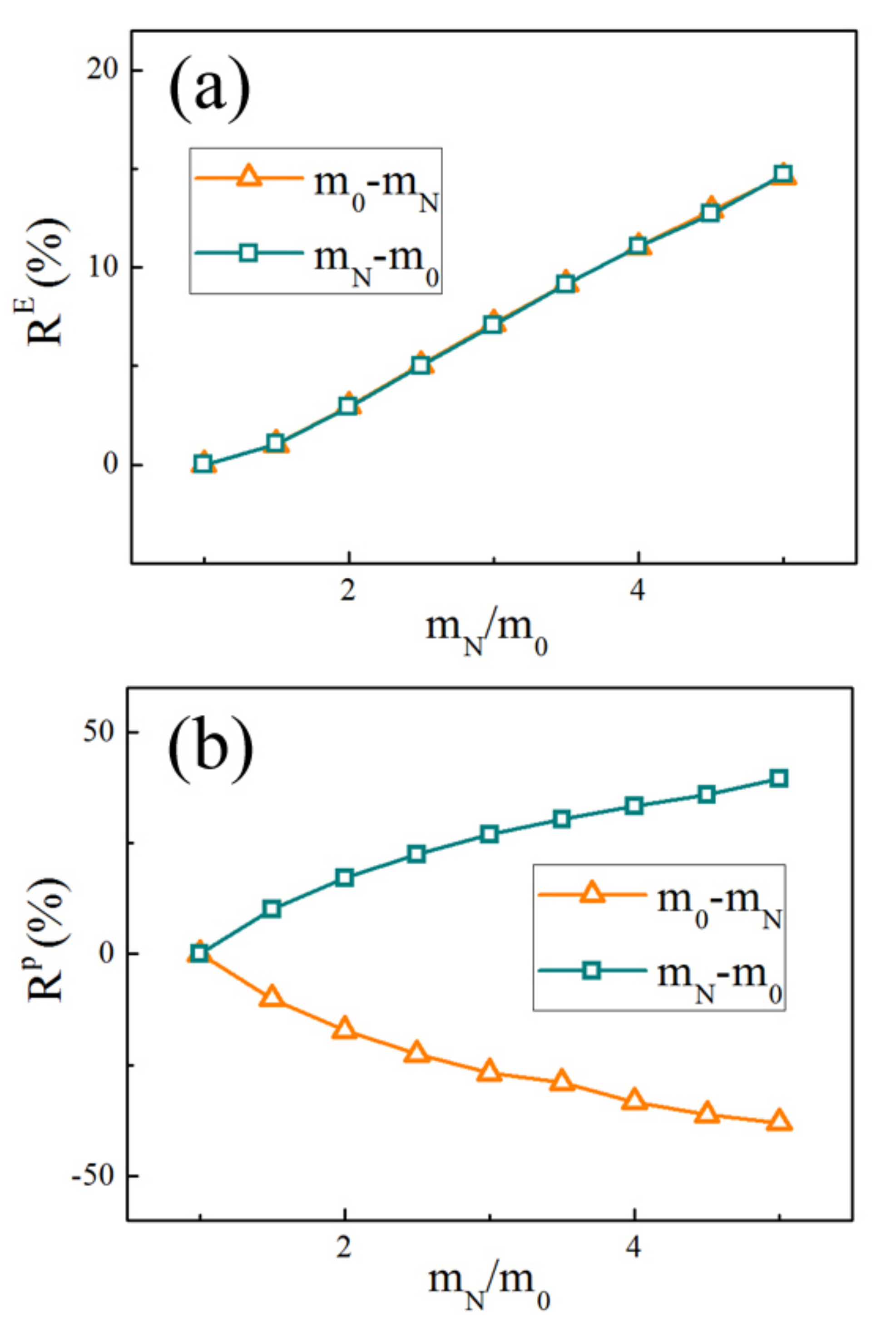}\caption{Here $m_{0}=12$ and $m_{N}$ varies from 12 to 60. For $m_{N}/m_{0}=1$,
it stands for the case that there is no mass interface in the graphene
flake. Thus the energy and the momentum reflection rates are both
0 in this case. (a) Energy reflection rate $R^{E}$ vs mass difference
ratio $m_{N}/m_{0}$. (b) Momentum reflection rate $R^{P}$ vs mass
difference ratio $m_{N}/m_{0}$.}

\end{figure}

Here we point out that the role of the KdV solitons surmised by Chang
et al. plays in thermal rectification is still under heavy debate\citet{01.Casati.Nature,12.cnt1,13.cnt2,14.cnt3,37.sufficient}.
First, an extremely large excitation is needed to generate the supersonic
KdV solitons. In carbon nanotubes, they are surmised to be generated
by the collision of electrons, ultrashort laser light or strong compressions\citet{38.cnt kdv,39.cnt kdv2}.
So it is almost impossible that in the heat conduction process at
room temperature, the KdV solitons could be generated. So far there
is still no direct evidence of the supersonic KdV solitons in either
carbon nanotubes or graphene flakes. Second, Chang et al. suggest
the preferred direction of the heat flux is from the heavy to the
light mass regions by considering KdV solitons\citet{11.science}.
However in the MD simulations of carbon nanotubes, it shows that the
preferred direction is from the light to the heavy mass regions. It
contradicts the existence of the KdV solitons in carbon nanotubes\citet{12.cnt1,13.cnt2,14.cnt3}.
In our molecular dynamics simulations of graphene flakes, there is
no thermal rectification effect in the heat conduction process. The
result is also against the existence of the KdV solitons in graphene.
Third, the square of the amplitudes is used to estimate the energy
of a soliton by Chang et al. However, the dependence of energy upon
the amplitude is far more complicated\citet{35.zhao,36.zhao2}. Thus
the width of a soliton is also necessary to be taken into consideration
in order to measure the amount of energy.

In summary, the inhomogeneous mass interface cannot bring thermal
rectification effect in graphene. The absence of rectification effect
is confirmed by studying different heat flux and mass difference ratio.
The microscopic mechanism is explained by the interfacial scattering
of solitons in graphene which reveals directional independent energy
reflection rate. Our results imply that the mass interface or the
mass gradient which is a combination of multiple mass interfaces cannot
be applied to design graphene based thermal rectifiers.
\begin{acknowledgments}
We thank Jiao Wang and Yong Zhang for helpful discussion and preparing
of the manuscript. This work was supported by National Natural Science
Foundation of China(\#10775115 and \#10925525).\end{acknowledgments}

\end{document}